# Demonstration of the Brightest Nano-size Gamma Source

A. S. Pirozhkov[1], A. Sagisaka[1], K. Ogura[1], E. A. Vishnyakov[2], A. N. Shatokhin[3], C. D. Armstrong[4], T. Zh. Esirkepov[1], B. Gonzalez Izquierdo[1,5], T. A. Pikuz[6], P. Hadjisolomou[2], M. A. Alkhimova[7], C. Arran[8], I. P. Tsygvintsev[9], P. Valenta[2], S. A. Pikuz[10], W. Yan[2,11,12], T. M. Jeong[2], S. Singh[13,14], O. Finke[2], G. Grittani[2], M. Nevrkla[2], C. M. Lazzarini[2,15], A. Velyhan[2], T. Hayakawa[1], Y. Fukuda[1], J. K. Koga[1], M. Ishino[1], Ko. Kondo[1], Y. Miyasaka[1], A. Kon[1], M. Nishikino[1], Y. V. Nosach[16], D. Khikhlukha[2], A. O. Kolesnikov[3], E. N. Ragozin[3], V. A. Gasilov[17], D. Kumar[2], J. Nejdl[2], P. V. Sasorov[2], S. Weber[2], D. Margarone[2], Y. Kato[18], G. Korn[2,5], H. Kiriyama[1], K. Kondo[1], C. Ridgers[8], T. Kawachi[1], M. Kando[1], and S. V. Bulanov[1,2]

e-mail: pirozhkov.alexander @ qst.go.jp

23 December 2024

1   Kansai Institute for Photon Science (KPSI), National Institutes for Quantum Science and Technology (QST), 8-1-7 Umemidai, Kizugawa, Kyoto 619-0215, Japan

2   ELI Beamlines Facility, The Extreme Light Infrastructure ERIC, Za Radnicí 835, Dolní Břežany 252 41, Czech Republic

3   Lebedev Physical Institute, Russian Academy of Sciences, Moscow, 119991 Russia

4   Central Laser Facility, Rutherford Appleton Laboratory, STFC, Chilton, Didcot, Oxon OX11 0QX, United Kingdom

5   Marvel Fusion GmbH, Theresienhöhe 12, Munich 80339, Germany

6   Institute for Open and Transdisciplinary Research Initiatives, Osaka University, Suita, Osaka 565-0871, Japan

7   Joint Institute for High Temperatures, Russian Academy of Sciences, Moscow 125412, Russia

8   York Plasma Institute, University of York, Heslington, YO10 5DQ, UK

9   ISTEQ AR, Raffi Street 111, Yerevan, Armenia

10  HB11 Energy Holdings, Freshwater, NSW 2095, Australia

11  Key Laboratory for Laser Plasmas, Shanghai Jiao Tong University, Shanghai 200240, China

12  Collaborative Innovation Center of IFSA, Shanghai Jiao Tong University, Shanghai 200240, China

13  Institute of Plasma Physics ASCR, Za Slovankou 1782/3, 182 00 Prague, Czech Republic

14  FZU – Institute of Physics ASCR, Na Slovance 1999/2, 182 21 Prague, Czech Republic

15  Czech Technical University in Prague, FNSPE, Brehova 7, 11519 Prague, Czech Republic

16  Institute of Physics, National Academy of Sciences of Ukraine, Prosp. Nauky 46, 03680 Kyiv, Ukraine

17  Keldysh Institute of Applied Mathematics, Russian Academy of Sciences, Moscow 125047, Russia

18  Institute of Laser Engineering, Osaka University, 2-6 Yamadaoka, Suita, 565-0871 Osaka, Japan

**Gamma rays selectively interact with nuclei, induce and mediate nuclear reactions and elementary particle interactions, and exceed x-rays in penetrating power and thus are indispensable for analysis and modification of dense objects. Yet, the available gamma sources lack sufficient power and brightness. The predicted and highly desirable laser-driven gamma flash, from here on termed "Gamma Flash", based on inverse Compton scattering from solid targets at extreme irradiances ($>10^{23}$ W/cm$^2$), would be the highest-power and the brightest terrestrial gamma source with a 30-40% laser-to-gamma energy conversion. However, Gamma Flash remains inaccessible experimentally due to the Bremsstrahlung background. Here we experimentally demonstrate a new interaction regime at the highest effective irradiance where Gamma Flash scaled quickly with the laser power and produced several times the number of Bremsstrahlung photons. Simulations revealed an attosecond, Terawatt Gamma Flash with a nanometre source size achieving a record brightness exceeding $10^{23}$ photons/mm$^2$mrad$^2$s per 0.1% bandwidth at tens of MeV photon energies, surpassing astrophysical Gamma Ray Bursts. These findings could revolutionize inertial fusion energy by enabling unprecedented sub-micrometre/femtosecond resolution radiography of fuel mixing instabilities in extremely-compressed targets. The new gamma source could facilitate significant advances in time-resolved nuclear physics, homeland security, nuclear waste management and non-proliferation, while opening possibilities for spatially-coherent gamma rays.**

## Introduction

The elementary processes of Gamma Flash are nonlinear Thomson and inverse Compton scattering[1,2]. They have been observed experimentally in collisions of intense laser pulses with electron beams from either conventional[3] or laser accelerators[4,5]. In either case, each interaction act involved single electron with multiple photons (up to $n$=4 in Ref.[3], $n$~500 in Ref.[4], and n~420 in Ref.[5]) resulting in the emission of a single gamma photon per electron. The number of interactions was relatively small (e.g. up to $N_\gamma$~$10^8$ in Ref.[5]) due to the low electron density in the pulse-beam collision geometry, thus limiting the achievable conversion efficiency. In contrast to these low particle number experiments, predicted by simulations[6,7] Gamma Flash occurs in the interaction of a high-irradiance laser pulse with solid-density target, **Fig. 1a**, where the electron density exceeds $10^{23}$ cm$^{-3}$, many orders of magnitude higher than achievable in beam experiments. The high-energy electrons are produced at-focus by the laser itself. Thus, similar to the theories of astrophysical Gamma Ray Bursts[8,9], high-energy photons in Gamma Flash are generated by relativistic electrons driven by and interacting with a strong electromagnetic field. Apart from the unprecedented 30-40% laser-to-gamma conversion efficiency, Gamma Flash is expected to produce a multi-Petawatt femtosecond pulse of MeV photons originating from a micrometre volume[6,7,10,11] and thus would have unprecedented brightness. The Gamma Flash regime would also be attractive due to the practicality of scheme allowing high repetition rates: a single laser beam and a simple solid target are envisioned, rather than alignment-sensitive multi-beam experiments or nanostructured and intermediate-density (between solid and gas) targets, as in some other proposed regimes reviewed in[12].

However, the generation of Gamma Flash is still challenging experimentally as it requires conditions where a solid-density material with the electron number density of >$10^{23}$ cm$^{-3}$ would be exposed to a very high irradiance, ~$10^{23}$ W/cm$^2$, which is just at the edge of present laser technology[13]. Due to laser imperfections, the actual at-focus irradiance in the experiments can be lower than expected[14]. In addition, target positioning errors, which usually remain unnoticed, prevent achieving high on-target irradiance[15], in which direct measurement is now arguably impossible. Further, the Gamma Flash conditions entail generation of a large amount of fast (MeV and GeV energy) electrons[16,17] going through the target and producing Bremsstrahlung[16,18-21]. Bremsstrahlung is widely seen even at much lower laser irradiances and originates from much larger (sub-millimetre) source during much longer (picosecond-scale) duration[22,23] and, therefore, has orders-of-magnitude lower brightness; however present gamma detectors have insufficient spatial and temporal resolution[24,25] to measure the difference. Further, Bremsstrahlung typically exhibits hard-to-distinguish spectral and angular distributions, **Extended Data Fig. 1a,b**, while the large scales involved make accurate simulation of its conversion efficiency difficult[23]; experimentally, a few percent conversion efficiency into Bremsstrahlung photons is demonstrated[20]. Therefore, just detection of a large number of MeV photons at presumably high, not directly measured, laser irradiance cannot be a decisive proof of the Gamma Flash occurrence, unless it is differentiated from Bremsstrahlung. Thus, till now, despite being the most important goal for many advanced laser facilities[26], Gamma Flash has not been demonstrated.

## Proposed method

In general, when a Petawatt-class[26] laser interacts with a solid target, its relatively weak nanosecond-long leading part (*prepulse*)[17] can erode a portion of the target making a so called preplasma with a typical exponential-like density profile. It is this type of the density profile that is advantageous for Gamma Flash[7,10,11]. The best choice of the preplasma scalelength depends on laser parameters, but is typically from a few to ten µm. The most intense femtosecond part (*main pulse*) of the laser beam is partially absorbed in the preplasma and is partially reflected from the remaining dense target, **Fig. 1b**. If the main pulse does not propagate deep inside the target, Gamma Flash occurs only in the preplasma, **Fig. 1a**, where both the laser-induced field and laser-accelerated electron energies are high enough. Therefore, its photon yield is independent of the target thickness, as observed in simulations[10], **Fig. 1c**. On the contrary, Bremsstrahlung originates from fast electrons propagating through the target bulk and, for a fixed electron population, its yield is proportional to the ion charge squared and material's thickness (for micrometre scale), as seen in simulations[18,22]. Based on this difference, we set up an experiment to measure the dependence of the hard-energy photon spectrum on the target thickness under fixed irradiation conditions, and then extrapolate the resulting linear dependence to zero thickness, thus revealing the pure Gamma Flash contribution, **Fig. 1c**.

Two intimately related requirements were crucial for the realization of this method. The first is *a high on-target laser irradiance* necessary for Gamma Flash. In our case the on-target irradiance was additionally increased due to three factors: self-focussing in the preplasma[27], a constructive interference with radiation reflected from the remaining target surface, and an optical focussing of the reflected radiation and its harmonics[28-31], **Fig. 1b**, due to the curvature of the remaining surface. These phenomena are seen in our 3D hydrodynamic and Particle-In-Cell (PIC) simulations, **Extended Data Fig. 2** and **Fig. 1a,b**. The second requirement was *a high level of laser-target interaction control* for obtaining *the same interaction for all target thicknesses* necessary for the sought-for linear dependence of the gamma photon yield on thickness. This included small shot-to-shot laser parameter variation, careful selection of the targets (Methods), and accurate target positioning to the laser focus[15]. To ensure the same interaction conditions, we kept the target material the same and varied the thickness only. The thinnest targets were thicker than the region eroded by the laser.

## Experimental setup and conditions

The experimental setup is shown in **Extended Data Fig. 3**. We used the J-KAREN-P laser[14,32] that produced a 32 femtosecond (Full Width at Half Maximum, FWHM) main pulse focused to the irradiance of $(4\pm2) \times 10^{21}$ W/cm² at the target surface in vacuum, corresponding to the dimensionless amplitude[17] $a_0 = eE_0/m_e c\omega$ of 43±10 (Methods); here $e$ and $m_e$ are the electron charge and mass, $c$ is the velocity of light in vacuum, and $E_0$ and $\omega$ are the laser field and angular frequency. According to simulations[10], such a pulse is capable of producing Gamma Flash, although with a smaller laser-to-gamma conversion efficiency compared to the theoretically proposed cases with $>10^{23}$ W/cm² irradiance. As discussed above and shown by the simulations, the actual in-plasma irradiance was much higher: the laser field energy density was enhanced by two orders of magnitude, **Fig. 1b**. We used stainless steel tapes of 5, 10, 15, and 20 μm thickness moved between shots, keeping the laser focus on the target surface[15]. The tape target was selected due to the high shot-to-shot reproducibility and high repetition rate allowing hundreds of shots in a day. Stainless steel was selected due to: its high mechanical strength, sufficient to keep the tape from breaking; practical absence of deformation (in particular, thinning) from tension keeping the tape flat; high, mirror-like, surface quality; and not-so-high Z-number of composing ions.

To tailor the preplasma and achieve the required scalelength of a few-μm, we used a controllable amount of sub-nanosecond prepulse preceding the femtosecond main pulse. In the experiment we observed that a thicker preplasma could increase the high-energy photon yield, **Extended Data Fig. 4i**, due to stronger self-focussing and greater laser absorption. However, it also increased shot-to-shot fluctuations and the risk of the tape breaking. A signature of the preplasma and interaction properties is the *surface phenomenon* of laser reflection including harmonics generation, because it occurs at a certain density and scalelength of preplasma[33,34] and is strongly irradiance-dependent[15,35]. We measured reflection at two different angles and specularly-reflected low-order harmonics for all the thicknesses, **Fig. 1d**. We found the conditions when the preplasma and interaction were the same for all the thicknesses, including the thinnest targets, confirming sufficient minimum thickness.

As high-energy (~0.1 to tens of MeV) photon detectors, we used two absolutely-calibrated linear absorption spectrometers (Methods), each consisting of 8 scintillators, **Fig. 1e**. One detector was placed in the forward (laser propagation) direction, and another approximately along the electric field polarization direction, 98° from the laser axis.

## Results

We observed linear dependences of the photon yield with respect to the target thickness, **Fig. 1f**, similar to the expected dependences, **Fig. 1b**. This was seen in all sixteen scintillators, eight of the forward and eight of the side spectrometers, **Extended Data Fig. 4** and **Extended Data Fig. 5**. The extrapolation of these linear dependences to zero thickness gave the Gamma Flash yield, i.e. eight values for each of the two spectrometers, upper circles in **Fig. 2a,e**. The eight slopes of these linear dependences revealed the Bremsstrahlung contributions (which are proportional to thickness); this contribution for 5 μm targets is shown by the lower rhombi in **Fig. 2a,e**. The eight scintillator signals from each spectrometer, **Fig. 2a,e**, allowed estimating the spectra of Gamma Flash and Bremsstrahlung in both observation directions assuming a power-law with exponential cutoff spectral shape with three parameters: number of photons $N_p$, power-law exponent $p$ and temperature $T$: $dN/d\mathcal{E} = (N_p/T)(T/\mathcal{E})^p \exp(-\mathcal{E}/T)$ (Methods). **Fig. 2b,f** show the parameter maps; **Fig. 2c,d,g,h** show the absolute spectra.

Notably, for the thinnest 5 μm target, Gamma Flash dominated over Bremsstrahlung, which is important for research and applications. For example, the dependence of the scintillator #6 signal of the forward spectrometer on the peak irradiance estimated from the measured on-shot energy and far field data is shown in **Fig. 3a**. This dependence was obtained with the 5 μm target, for which Gamma Flash dominated, and thus we conclude that Gamma Flash was brighter at higher irradiance. On the other hand, for the thick (20 μm) target, for which Bremsstrahlung was on average stronger, the dependence on irradiance was apparently absent, **Fig. 3c**. Assuming that the Bremsstrahlung contribution weakly depends on irradiance, we subtracted its average contribution from the data, thus obtaining the dependences of the Gamma Flash contribution on peak irradiance, shown in Log-Log scale in **Fig. 3b,d**. Power-law fits of these data gave the same, within the error bars, dependences. However, the error bars from the 5 μm thick targets were smaller due to smaller impact of the Bremsstrahlung background – the advantage of the Gamma Flash dominance. The laser pulse energy and power were approximately constant during all the shots, **Extended Data Fig. 6c,d**; the irradiance difference was caused by shot-to-shot spot quality variations. The resulting $\sim I^{0.5 \pm 0.3}$ dependence indicates that even for the same laser pulse power, the dependence on the spot quality is significant and must be considered in research and applications.

We obtained the Gamma Flash yield dependence on the peak laser power by repeating the experiment at a lower power of $P = 122 \pm 6$ TW (~44% of the full $P = 280 \pm 20$ TW power), **Fig. 3e,f**. For relatively soft photon energies, 0.1 MeV and above, the yield depended as $\propto P^{1.9\pm0.2}$. With increasing photon energy, the dependence became steeper, and for photons above 10 MeV the yield grew as $\propto P^{3.1\pm0.8}$, indicating favourable scaling. In contrast to the full power experiment, at the reduced laser power Gamma Flash was not dominant even with the thinnest 5 μm targets, **Fig. 3e**.

Gama Flash spectra obtained from 3D PIC simulation (Methods) were in good agreement with the experimental data for both observation directions, **Fig. 2c**,**g**. The full-angle Gamma Flash spectrum and conversion efficiency obtained from 3D simulation are shown in **Extended Data Fig. 2f** and **g**, respectively. The conversion efficiency to photons with energy above 0.1 MeV was ~0.3%. As a 3D simulation of Bremsstrahlung is still out of reach due to the significant volume and timescale involved, we provide a comparison of the normalized experimental spectral shapes with a 2D simulation in **Extended Data Fig. 1d**, which also shows good agreement.

Although even state-of-the-art methods do not allow measurements of the Gamma Flash source size and pulse duration yet, they can be retrieved from simulations. The full source size and duration in our 3D simulation were ~1 μm and 20 fs, respectively, **Fig. 1a**; the gamma photon number was $1.2\times10^{11}$ and the total power was 1.1 Terawatt, from which 1 Terawatt was provided by photons above 1 MeV, **Extended Data Fig. 2c**. Further, for applications important measures are size and duration corresponding to an observation direction and acceptance angle of a diagnostics or sample. For such a case of narrow-angle forward emission, our simulations revealed that the Gamma Flash source size was much smaller and the duration was much shorter, both quickly decreasing with photon energy. **Fig. 4a-d** show sources of the forward emission in several photon energy ranges; **Fig. 4e**,**f** show rms source size and rms duration dependences on photon energy and their exponential fits. At high photon energies, the rms source size became smaller than 100 nm and the rms duration became close to 1 fs, the values much smaller than the driver laser pulse. Moreover, the forward Gamma Flash at high photon energies was emitted as an attosecond pulse train, **Fig. 4g**.

## Discussion

High photon yield, small source size and short duration make the Gamma Flash brightness unprecedented in the MeV and tens of MeV photon energy range, including the giant dipole resonance (GDR) spectral region. Comparison with other experimental results is shown in **Fig. 4h**. Moreover, the brightness was comparable to the case of a Gamma Flash driven by a 10 PW laser in simulations[11]. Favourable laser irradiance and power scalings, **Fig. 3**, promise extension to higher brightness and photon energies.

Attosecond pulses[36] are commonly generated in the extreme ultraviolet (several tens to a few hundred eV) spectral region, and have only recently become available at 9 keV[37], i.e. at three orders of magnitude lower photon energies than the presented gamma source. An attosecond gamma pulse train is predicted by simulations of two-pulse interaction with a microwire[38]; in contrast, a single pulse and simple tape targets were employed in the presented regime.

The source was small and the duration was short because simultaneous high irradiance and high electron energy density could happen only in a small volume where the reflected intensified field overlapped with the electron density spike, **Fig. 1b**, and where the simulation showed the quantum nonlinearity parameter $\chi_e$ values up to 0.23, **Extended Data Fig. 2e**, with an estimated nonlinearity order $n$~$10^6$ to $10^9$ (Methods). This overlap happened each laser period near the laser pulse peak, Movie 1, also evidenced by the periodic attosecond pulse train, **Fig. 4g**. The laser-electron interaction regime achieved in the overlap region (filled star in **Fig. 5**) corresponding to the maximum $\chi_e$ and irradiance was very different from the regime corresponding to the averaged $\chi_e$ and original laser irradiance in the entire focal volume (open star in **Fig. 5**): the former corresponds to radiation-dominated dynamics[39,40]. The high field intensity was ensured by the laser pulse self-focussing in the preplasma[27] and focussing of the harmonics reflected from the concave surface prepared by the prepulse. The latter effect was similar to harmonics focussing due to the target surface denting by the main pulse itself[28-31], however, at our irradiance and density this denting was much weaker and therefore the irradiance increase due to this effect was small. The presence of the concave reflecting surface in the experiment was evidenced by the generation of high-order harmonics with wavelengths down to at least 17 nm (spectrograph limited, shorter wavelengths were probably generated). The high density of the electron spike in the preplasma is explained by catastrophe theory[41], which states that structurally-stable fold singularities are always produced in multistream plasma flows, evidenced in our case by characteristic phase space structures and $\propto(\xi-\xi_0)^{-1/2}$ density dependence observed in the 3D PIC simulation, **Extended Data Fig. 2d** (inset) and Movie 2. This is similar to the Burst Intensification by Singularity Emitting Radiation (BISER) mechanism[42,43], although full coherence is probably not achieved in the Gamma Flash emission due to its orders-of-magnitude shorter wavelengths, in the picometre and sub-picometre range.

Nevertheless, similar to the Young's Double Slit Experiment, the small Gamma Flash source size provides conditions for the local spatial coherence, **Fig. 4i**. In contrast to low-brightness sources, Gamma Flash can provide significant photon numbers in a single femtosecond flash, **Fig. 4j**, alleviating the sub-atomic-scale setup stability requirement in potential gamma-interference and gamma-diffraction experiments. Thus, Gamma Flash may become the first practical coherent gamma source, instead of still-hypothetical gamma-ray lasers[44]. This opens a way to phase contrast imaging[45] and other spatial-coherence-required applications in the sub-MeV and eventually MeV photon energy ranges, either single-shot or at a ~1 Hz or higher repetition rate provided by already available PW-class lasers[26] with simple repetitive tape target drivers, as in our experiment.

The small gamma-ray source size and short pulse duration in our regime enable sub-micrometre and a few-femtosecond resolution radiography, which is inaccessible with other sources, such as Bremsstrahlung[24,25,46], due to their orders-of-magnitude larger source sizes and longer pulse durations. Micrometre/femtosecond-resolution radiography with highly penetrating gamma rays may show asymmetry and the actual state of burning fuel mixing instabilities of dense, extremely-compressed targets in the inertial fusion programme[47]. Asymmetry and Rayleigh–Taylor and Richtmyer–Meshkov instabilities are major research topics of interest as they limit the compression ratio and therefore the burning efficiency and energy gain[48]. Yet, they remain inaccessible till now with any experimental methods, while their simulations are model-dependent. Thus, our findings may lead to a breakthrough in inertial fusion energy.

The Gamma Flash experiment constitutes an example of process simulation for laboratory astrophysics[49,50], where physical conditions can be qualitatively compared to extraterrestrial regimes. Formation of periodically emerging slightly dephased caustics in electromagnetic energy density and electron density seen in 3D PIC simulation resembles the conditions that may occur near the surface of a rotating magnetar[51].

Our results experimentally proved the Gamma Flash concept, validated the simulation codes, and will become the basis for the following experiments at modern and future laser facilities, eventually opening applications in nuclear physics[52,53] (including time-resolved studies), homeland security[54,55], nuclear non-proliferation and radioactive waste management[56], high-resolution radiography of burning fuel in laser fusion[46,47], laboratory astrophysics[49,50], and coherent gamma ray optics.

## Methods

### Laser

The experiment was performed with the J-KAREN-P laser facility[32]. Like other state-of-the art high-power lasers[26], J-KAREN-P provides pulses consisting of the main high-power femtosecond pulse and very small amount of nanosecond-duration light (prepulse) preceding the main pulse with a typical power contrast of ~$10^{-10}$ or even smaller[57,58]. We used a so-called broadband mode with larger bandwidth, resulting in shorter pulses, higher power, and somewhat lower contrast ($10^{-10}$ in our case) compared to the so-called high-contrast mode used in other high-irradiance experiments with the J-KAREN-P laser[59-61]. The lower contrast resulted in modification of the target surface, namely, by generation of the small-scale preplasma. By controlling timing and energy of the pumping lasers, we fine-tuned the contrast to achieve the benefits stated in the main text – namely, optimum self-focussing increasing the laser irradiance, higher laser absorption, and most stable Gamma Flash generation. The preplasma also weakened the electrostatic sheath potential thus reducing electron reflection at the target surfaces (recirculation[62,63]), which could change the expected linearity of Bremsstrahlung at small thicknesses. We note that in our experiments attempted at lower contrasts (and therefore larger-scale preplasma) the gamma-ray yield was sometimes higher, however, its stability was lower (shot-to-shot fluctuations were larger), **Extended Data Fig. 4i**, which decreased the accuracy of the Gamma Flash differentiation from Bremsstrahlung. Also, the 5 µm tapes were often broken in the lower-contrast experiments. In the presented regime the gamma yield was close to optimum and its stability was high, **Fig. 1f**, which allowed the Gamma Flash demonstration.

The laser parameters were measured carefully for understanding the interaction physics and correct setting of simulation parameters. The pulses had a central wavelength of $\lambda_0$ = 804.9±0.9 nm (here and below mean ± shot-to-shot standard deviation) and a pulse energy of $E_L$ = 9.5±0.5 J. In the reduced power mode, the pulse energy was $E_L$ = 4.16±0.16 J. Both the central wavelength and pulse energy were determined from the absolutely calibrated full-power on-shot spectra, because this method provided the highest accuracy and lowest shot-to-shot fluctuations among several methods we used[15]. The off-axis parabola became gradually contaminated due to the target debris; the corresponding losses of laser pulse energy were taken into account. During the entire experiment (1759 shots), the pulse energy reduced by 16%; during one experimental day presented here (342 shots), the losses were ~2.6%. Pulse energies during the data shots presented in this paper are shown in **Extended Data Fig. 6c**.

The focal spot, **Extended Data Fig. 6a**, was measured with two methods which gave fully consistent results: (i) measured at-focus in-vacuum spot with a microscope objective and CCD camera, with the laser beam attenuated with wedges, on the same day before shots at ~10% of the maximum laser power, which gave similar spot shapes as the full-power measurement as we confirmed in a separate experiment[14]; and (ii) measured full-power on-shot far field, the beam for this measurement was transmitted through a high-quality dielectric mirror and focused by an off-axis parabola[14].

The temporal laser pulse shape was also determined via two methods: (i) measured on the same day before shots with self-referenced spectral interferometry[64] (a commercial Wizzler[65]) at ~10% of the maximum laser power, which gave the same pulse duration as the full-power measurement as we confirmed in a separate experiment[66]; and (ii) calculated from the full-power on-shot laser spectra and the spectral phase from the Wizzler measurement. Both methods gave similar pulse shapes, **Extended Data Fig. 6b**.

The peak power was calculated from the pulse energy and pulse shape taking into account low-power wings as $P = E_L/\tau_{Eff}$, where $\tau_{Eff}$ = 34.5±0.7 fs is the effective pulse width, i.e. the area under the normalized power curve shown in **Extended Data Fig. 6b**: $\tau_{Eff} = \int p(t)dt$. Similarly, the vacuum irradiance was calculated from the peak power and focal spot normalized irradiance distribution $I_{norm}(x,y)$ shown in **Extended Data Fig. 6a** considering low-irradiance halo as $I = P/S_{Eff}$, where the effective spot area was $S_{Eff} = \pi r_{Eff}^2 = \int I_{norm}(x,y)dxdy$ with the effective radius $r_{Eff}$ = 1.5±0.2 µm. The contrast on the nanosecond and tens of picosecond time scales was measured at full power on-shot by a fast photodiode and high-bandwidth oscilloscope with a signal rise time of 30 ps.

## Setup

The p-polarized laser pulses were focused with an f/1.3 Off-Axis Parabolic (OAP) mirror with a 45° deviation angle onto the surface of a 20-mm wide tape target (described in the next section) at an incidence angle of 45°, **Extended Data Fig. 3**. This incidence angle was selected to ensure laser safety, as in normal-incidence cases often used in simulations the back-reflection could destroy the laser. The spectra of the reflected laser radiation and its low-order harmonics were measured with a fibre spectrometer on a PTFE screen situated at ~1.5 m from the target. The acceptance angle was 9 msr near the specular direction. The back-reflected radiation collected by the full acceptance angle (0.52 sr) of the main off-axis parabola was measured for diagnostics and laser safety purposes. Both setups, the fibre spectrometer + screen and the back-reflection diagnostics, were absolutely calibrated, thus providing energies of the back-reflected and specularly reflected 1st, 2nd, and 3rd harmonics, **Fig. 1d**. These data, being sensitive to the laser contrast and irradiance[21,67,68], were used to ensure identical conditions of the shots with different target thicknesses. **Fig. 1d** also shows the difference from the lower-contrast regime shown by the grey crosses; interestingly, observation of the laser frequency reflection (1ω) *only* would not allow reliable interaction regime monitoring, although the error bar of the lower-contrast mode in this particular diagnostic was larger.

**Targets**

Without tuning laser parameters and careful target selection and in-focus positioning, Gamma Flash remains either too weak (non-measurable) or indistinguishable from Bremsstrahlung. Fine laser tuning and correct target selection and positioning require hundreds or even thousands of shots. We therefore selected a tape target system which allowed use of the full laser repetition rate of 0.1 Hz resulting in typically several hundreds of full-power shots per day. The shot-to-shot stability of the laser and tape target allowed shots under highly reproducible conditions, which facilitated the experiment.

We aimed for practical tape target material and thickness so that contributions from both Gamma Flash and Bremsstrahlung could be reliably measured and differentiated, which is required for a demonstration experiment. For too-thin or low-Z or mass-limited targets, the Bremsstrahlung would not be measurable – which may be advantageous for applications, but not suitable for a demonstration experiment. Also, too thin tapes were broken by the laser, resulting in a premature stop of the shot series. On the other hand, with too-thick or high-Z targets, the Gamma Flash contribution would become indistinguishable, while the Bremsstrahlung contribution might exhibit nonlinear thickness dependence due to attenuation in the target[18]. Furthermore, the material should have mechanical properties which would allow using it as thin tapes (in contrast to aluminium, for example, which was easily broken). We finally selected stainless steel due to its high mechanical strength allowing experiments with down to 5 μm thickness, mirror-like surface quality increasing result reproducibility, and low deformation under tension of the tape driver – in particular, this preserved the thickness, unlike more easily stretchable plastic targets. The Z-number of the composing ions (mainly Fe, Cr and Ni) was not very high, so Bremsstrahlung was not very strong, but measurable, as required for its reliable differentiation from Gamma Flash.

We used typically 10-20 preliminary full-power shots to place the tape target at the best focus with typically ~5-10 μm accuracy (the method is described in detail in[15]) and immediately after that took the data shots. Tapes of several thicknesses were preloaded into the target system in advance; all four target thicknesses, i.e. 5, 10, 15, and 20 µm were irradiated within one day without opening the target chamber with minimum time separation between the shot series.

**Linear absorption spectrometers**

Gamma rays from Gamma Flash and Bremsstrahlung were detected with two linear absorption spectrometers[69,70] situated behind the target in the forward (0°, along the laser) and side (98°, nearly along the laser field) directions. The side spectrometer was placed at this angle to sample the angular distribution, so that two very different spectra could be compared with the simulations: this allowed avoiding accidental similarity which might happen if only a single direction was used. It also helped to distinguish the Gamma Flash from other mechanisms with narrow angular distribution such as betatron radiation and other regimes in the expanded targets[71,72]. Both spectrometers were heavily shielded with lead blocks and lead and tungsten collimators. Permanent magnets were employed to divert electrons and ions. Radiation from the vacuum target chamber passed through a 2.5 mm Be, 75 µm Kapton window, and alignment Image Plate to reach the forward spectrometer, and a 3 mm stainless steel flange to reach the side spectrometer. The losses and secondary particle generation in these components were included into the spectrometer response obtained with GEANT4 (Ref. [73]) simulations, **Extended Data Fig. 3b,c**.

Each spectrometer consisted of eight 2-mm thick, 10 mm × 30 mm LYSO ($Lu_{1.8}Y_{0.2}SiO_5$) scintillators, with 2-mm-thick tungsten absorption filters after scintillators #4 to #7 (the numeration starts from #0), **Extended Data Fig. 3** inset. The fluorescence in the ~400-500 nm spectral range emitted from the scintillator array was imaged onto a CMOS camera; an example of raw data is shown in **Fig. 1e**. The beamline absorption and scintillator response allowed registering photons with energies above ~20 keV (forward) and ~50 keV (side spectrometer), **Extended Data Fig. 3b,c**. The harder radiation was detected by deeper scintillators with higher numbers in the array, however, there was no one-to-one correspondence, evidenced by the broad response curves in **Extended Data Fig. 3**. The absolute calibration and spectrum reconstruction method is described in detail in[70,74].

We used a spectral shape inspired by simulations[11,22], i.e. power-law with exponential cutoff with three parameters: number of photons $N_p$, power-law exponent $p$ and temperature $T$: $dN/d\mathcal{E} = (N_p/T)(T/\mathcal{E})^p \exp(-\mathcal{E}/T)$. The parameter values providing the best match to the experimental data are shown by the yellow circles in **Fig. 2b,f**. These figures also show the maps of the Bayesian Information Criterion (BIC)[75]; each point corresponds to a pair of values (temperature, power-law exponent) ($T$, $p$) for which the optimum number of photons $N_p$ was calculated with the analytical method[70]. The colour of each point corresponds to the Bayesian Information Criterion (BIC) difference, ΔBIC, from 0 (the best fit) to 2; for ΔBIC = 2, the parameters are significantly worse for fitting the given experimental data (1/e probability factor). Thus, the coloured regions provide estimates of the viable parameter volume. The absolute spectra with error bars estimated from the coloured regions are shown in **Fig. 2c**,**g** (Gamma Flash) and **d**, **h** (Bremsstrahlung). Comparison of the forward Gamma Flash and Bremsstrahlung spectra is shown in **Fig. 3e**.

## Software

We used Origin Pro[76] software with default settings for fitting the data from the experiment and simulations.

We used Mathematica[77] software and its FindMinimum[] function for implementation of the spectrum reconstruction method discussed in the previous section.

## Hydrodynamic Simulations

The preplasma formation occurs on time scales of ~1 ns at irradiances ~$10^9 - 10^{15}$ W/cm². Under these conditions, the characteristic time of the plasma dynamics is much higher than the times of thermalisation and ionisation equilibrium establishment, which allowed us to assume that local plasma properties (pressure, internal energy, ionic composition, transport coefficients, etc.) are completely defined only by its temperature and density, and use the hydrodynamic approach. Simulations of the preplasma, **Extended Data Fig. 2a**, were carried out with the 3DLINE code[78,79] designed for nanosecond-scale laser target dynamics and radiation simulations. The stainless steel equation of state was obtained using the FEOS code[80] based on the Thomas-Fermi average atom model with a quasi-empirical correction in the low temperature region to account for the liquid-gas phase transition. The laser energy transport and deposition were calculated using a hybrid model[81], which combines geometric optics in the low-density regions with the solution of the Helmholtz equation along rays near and behind the critical surface, the main laser absorption mechanism being inverse Bremsstrahlung. The emission and transport of the thermal plasma radiation were calculated using the multigroup diffusion approximation; the averaged emission and opacity coefficients were calculated with the THERMOS code[82] under the assumption of local thermodynamic equilibrium. The model also took into account the thermal conductivity in the plasma, which is mainly determined by the contribution of electrons. Details of the approach are described in[83].

## 2D Hybrid-PIC simulations

Hybrid-EPOCH[23,84,85] is an extension of the open-source particle-in-cell code EPOCH[86], which has been designed to calculate the rate of Bremsstrahlung emission over much longer timescales (10s-100s of picoseconds) than generally possible in standard PIC codes. Whereas EPOCH tracks all particles in the simulation, both electrons and background ions, and calculates the electro-magnetic fields self-consistently, Hybrid-EPOCH tracks only the fast electrons that contribute to Bremsstrahlung as they pass through a static solid background. It uses a resistivity model for electron transport[87] with a return current of background electrons[88] and TNSA boundary conditions which allow the fastest electrons to escape while refluxing most of the electron population[89]. Routines for tracking high energy electrons passing through matter and calculating Bremsstrahlung are taken from Geant4 (Refs. [73,90]).

The 2D Hybrid-EPOCH simulations were conducted using a fast electron population injected from 2D EPOCH simulations. The PIC simulation calculated both Inverse Compton Scattering (Gamma Flash) and Bremsstrahlung radiation for photons above 100 keV over the first 200 fs of the interaction. The fast electrons above 100 keV and the electric and magnetic field structures were then extracted from this simulation and used to run a Hybrid-EPOCH simulation, which calculated the Bremsstrahlung radiation caused by these electrons over a further 1.2 ps. In the PIC simulation, a linearly-polarised laser pulse was focussed onto an iron target at 45 degrees, reaching a peak irradiance of $5 \times 10^{21}$ W/cm$^2$ over a spot size of 1.83 μm (FWHM) and a Gaussian temporal profile with a FWHM duration of 32 fs. The target was modelled using a density profile extracted from the hydrodynamic simulation of the laser pre-pulse, with a 5 μm layer of fully ionised iron at an atomic density of $8.5 \times 10^{22}$ cm$^{-3}$. The PIC simulation covered a 30 μm × 30 μm region with a cell size of 2.5 nm and two quasiparticles per cell (600 million quasiparticles in total). The hybrid simulation used the same atomic density of iron over the same region but with a much larger cell size of 60 nm and with 190 million quasiparticles.

As 2D simulations assume unlimited extension of the box in one direction, the absolute spectra for the comparison with experiment can only be calculated with some arbitrary-selected assumption; we therefore show the normalized spectral shapes, **Extended Data Fig. 1c**,**d**.

## 3D PIC simulations

We performed the 3D PIC simulations via the relativistic quantum electrodynamic code EPOCH[86]. The preprocessor derivatives for Higuera-Cary[91] and quantum electrodynamics (QED)[92] were enabled for the code compilation. EPOCH QED assumes the condition $\alpha\chi_e^{2/3} < 1$ is satisfied[93] (where $\alpha$ is the fine structure constant), which was valid for the electron energies and fields present in our experiment, **Extended Data Fig. 2e** and **Fig. 5**. We switched off photon dynamics to save computational time, as at this irradiance the Breit-Wheeler pair creation mechanism can be neglected; instead, we saved the photon generation time to find the Gamma Flash duration. Photons with energies below 0.1 MeV were ignored.

The simulation box extended from −10.24 μm to 2.048 μm in the x-direction, from −10.24 μm to 10.24 μm in the y-direction and from −5.12 μm to 5.12 μm in the z-direction, where (0, 0, 0) was the laser focal spot location. The cell size was 4 nm × 4 nm × 32 nm. The small cell size was necessary to resolve the relativistically-corrected skin depth with a resolution of approximately 5 cells, for the target density and irradiance mentioned below. The number of quasiparticles, 2.3×10$^{10}$ for electrons and ions each, resulted in 4 quasiparticles per cell. The initial number density for the electrons and ions was imported into EPOCH from the hydrodynamic simulation.

The lower x- and y-boundaries were set to 'simple laser', with the rest of the boundaries set to 'open'. Two identical lasers were launched from the two laser-emitting boundaries (x- and y-), appearing as one laser emitted from the corner of the simulation box. The laser pulse (Gaussian) duration at FWHM was 32 fs with a focal spot of 1.8 μm × 1.6 μm in the polarisation plane and perpendicular direction, respectively. The incidence angle on the target was 45°, the laser was p-polarized. The peak irradiance at the (vacuum) focal spot was 5×10$^{21}$ W/cm$^2$. The laser emission was initiated at a time *t* = 0 corresponding to two standard deviations before the irradiance peak. The QED module was started with a delay of 50 fs, somewhat before the intense laser pulse interaction with the plasma (the Gamma Flash emission started at ~55 fs, **Extended Data Fig. 2c**).

## Brightness estimates

Brightness is calculated as *B* = $N_{0.1\%}$/τ*S*Ω, where $N_{0.1\%}$ is the photon number in 0.1% bandwidth and solid angle Ω, τ is the pulse duration, and *S* is the source area. Nanometre-scale gamma source size and femtosecond-scale gamma pulse duration cannot be measured even via state-of-the-art methods. We used rms pulse duration and source size for the conservative Gamma Flash brightness estimate from the 3D simulation; alternative measures, such as FWHM or effective duration and source sizes, would give much higher brightness: for example, the FWHM gamma pulse duration at ~20 MeV was 240 as, **Fig. 4g**, more than an order of magnitude shorter than the rms duration of ~4-5 fs, **Fig. 4f**.

For the Gamma Ray Burst (GRB) brightness, we used absolute spectrum of the "brightest of all time" GRB 221009A, the highest curve in Fig. 2B of Ref.[94]. For the GRB source size we used the standard estimate based on the GRB duration. The dashed curve in **Fig. 4h** was obtained with a 10 light second source size corresponding to a ~10 s duration of the strongest pulse of the GRB 221009A time profile[95]; an estimate based on the ~600 s full duration of this GRB would give 3,600 times lower brightness. The dotted line in **Fig. 4h** was obtained with the same spectrum of "the brightest of all time" long GRB 221009A and a 0.3 light second source size corresponding to the average duration of short GRBs[96], thus, representing a maximum GRB brightness estimate.

**Nonlinearity order estimate**

The nonlinearity order $n$, i.e. the number of optical photons interacting with a single electron during the emission of one gamma photon, for a monochromatic electromagnetic wave can be estimated[5] from the maximum gamma-photon energy $E_{\gamma max}$, maximum electron energy $\gamma m_e c^2$, and field dimensionless amplitude $a_0$:

$$E_{\gamma \max} = \frac{4\gamma^2 \hbar \omega\, n}{1 + \frac{a_0^2}{2} + 4\gamma \frac{\hbar \omega}{m_e c^2} n}.$$

Here we assumed counter-propagation; for other angles, the required nonlinearity order is higher.

In our case the focussed reflected electromagnetic pulse contained not only the original laser frequency $\omega$, but also its harmonics $m\omega$, where $m$ is the harmonic order, and in a single gamma photon emission event, optical photons of different frequencies were scattered together. Thus, the above expression can be used to find the minimum and maximum estimates of the nonlinearity order for the maximum $m_{max}$ and minimum ($m_{min}=1$) harmonic orders, respectively. The maximum effective harmonic order was $m_{max}$~5 (wavelength $\lambda_5 \approx 160$ nm), which can be estimated from the gamma source size at high photon energies, ~16 nm, Fig. 4e, and the minimum focal spot size of a sharply focused electromagnetic wave, ~0.1$\lambda$, Ref.[97] For the laser wavelength ($m_{min}=1$), the peak electric field of $2\times 10^{13}$ V/cm (in the 3D simulation) corresponds to $a_0$~430, while for the 5th harmonic it corresponds to $a_0$~86. With the maximum electron energy in the simulation of 120

MeV ($\gamma$=235), the gamma-ray spectrum could be extended up to ~100 MeV with the nonlinearity order between $10^6$ and $10^9$, **Extended Data Fig. 7**.

## Data availability

The data are available from the corresponding author on reasonable request.

## Code availability

The code used in this work is available from the corresponding author on reasonable request.

## Acknowledgements

We are grateful for support from the J-KAREN-P laser team and computational support from the University of York IT Services and the Research IT team. We acknowledge contribution from Professor David Neely who sadly passed away before this work was completed; he was a colleague, mentor, and friend to the authors. We used the Viking cluster, a high-performance compute facility provided by the University of York. Financial support was provided by JSPS KAKENHI Grant Numbers JP17F17811, JP19KK0355, JP19H00669 and JP22H01239, QST-IRI, QST President's Strategic Grant (Creative Research), MEXT Grant Number JPMXS0450300221, ELI-Beamlines, project Advanced Research using High Intensity Laser Produced Photons and Particles (ADONIS) (Project No. CZ.02.1.01/0.0/0.0/16_019/0000789) from the European Regional Development Fund, Ministry of Youth and Sports of the Czech Republic (Project Nos. LM2023068 and LM2018114), National Key R&D Program of China 2021YFA1601700, and Central Laser Facility's EPAC diagnostic work package.

## Author contributions

A.S.P., T.Zh.E., and S.V.B. conceived the research. A.S.P. conceived and led the experiment. Y.M., A.S.P., A.S., K.O., Y.F., Ko.K., A.K., K.K., M.K., and H.K. prepared and tuned the laser and beamline. A.S., K.O., E.A.V., A.N.S., B.G.I., T.A.P., P.H., M.A.A., S.A.P., W.Y., T.M.J., S.S., O.F., G.G., M.N., C.L., A.V., T.H., Y.F., M.I., Ko.K., A.K., Ma.N., J.N., A.O.K., E.N.R., M.K., and A.S.P. prepared and performed the experiment. Y.V.N. and E.A.V. estimated laser pulse energy loss due to the debris. C.D.A. designed, developed, and calibrated the linear absorption spectrometers. C.D.A., D.Ku., and A.S.P. performed the linear absorption spectrometers data processing and gamma spectra reconstruction. I.P.T. and V.A.G. performed hydrodynamic simulations. P.H., T.Zh.E., C.A., P.V., J.K.K., D.Kh., and C.R. performed hybrid-PIC and PIC simulations. T.Zh.E., P.V.S., C.R., and S.V.B. developed the theory. A.S.P. and T.Zh.E. investigated the laser-plasma interaction regime and gamma radiation source properties. A.S.P., S.W., D.M., Y.K., G.K., K.K., T.K., M.K., and S.V.B. supervised the research. A.S.P., P.H., C.D.A., C.A., I.P.T., J.K.K, C.R., S.V.B., and T.Zh.E. wrote the manuscript. All authors edited the manuscript and contributed to its final version.

**Competing interests**

The authors declare no competing interests.

**Additional information**

Supplementary information: Movie 1 and Movie 2

Corresponding author: A.S.P., e-mail: pirozhkov.alexander@qst.go.jp

**Figures**

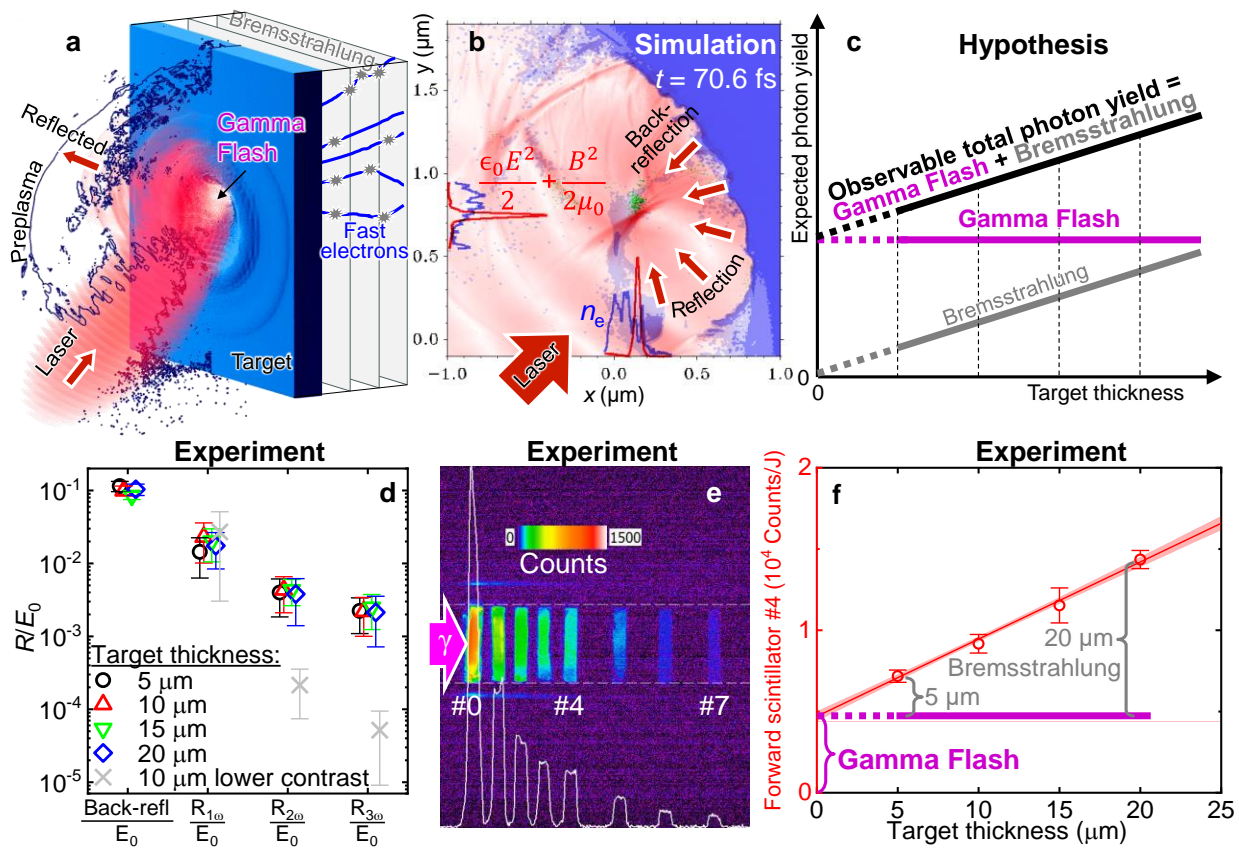

**Fig. 1. Gamma Flash demonstration. a**, 3D PIC simulation + sketch: the laser induces Gamma Flash (white) in the prepulse-created preplasma. Laser-driven fast electrons escaping the interaction region produce Bremsstrahlung (random grey stars in the target bulk) proportional to the varied-in-steps target thickness. **b**, 3D PIC simulation: the laser upon reflection from the dense plasma (solid blue) generates high-order harmonics focussed to a 104 times higher-than-initial peak energy density corresponding to an effective irradiance of $5.2\times10^{23}$ W/cm$^2$ (red colourscale and lineouts) and electron density peak (blue lineouts) reaching $3\times10^{23}$ cm$^{-3}$, **Extended Data Fig. 2d**; overlap of both peaks creates Gamma Flash photons (green). **c,** Idea: the expected Gamma Flash yield remains constant, while the Bremsstrahlung yield grows in proportion to the target thickness. Their sum is the total observable photon yield; its extrapolation to zero thickness gives Gamma Flash. **d**, Experiment: evidence of front-target-surface processes independence from the thickness: energies of back-reflection and the 1$^{st}$ to 3$^{rd}$ harmonics near the specular direction normalized by the laser energy $E_0$; error bars: standard deviations. **e**, Experiment: eight scintillators of the side spectrometer (5 µm target, 21-shot average). **f**, Experiment: high-energy photon signal dependence on the target thickness and its linear fit (line) and 67% confidence band (shaded); error bars: standard errors. **Extended Data Fig. 4** and **Extended Data Fig. 5** show data from all scintillators of both spectrometers.

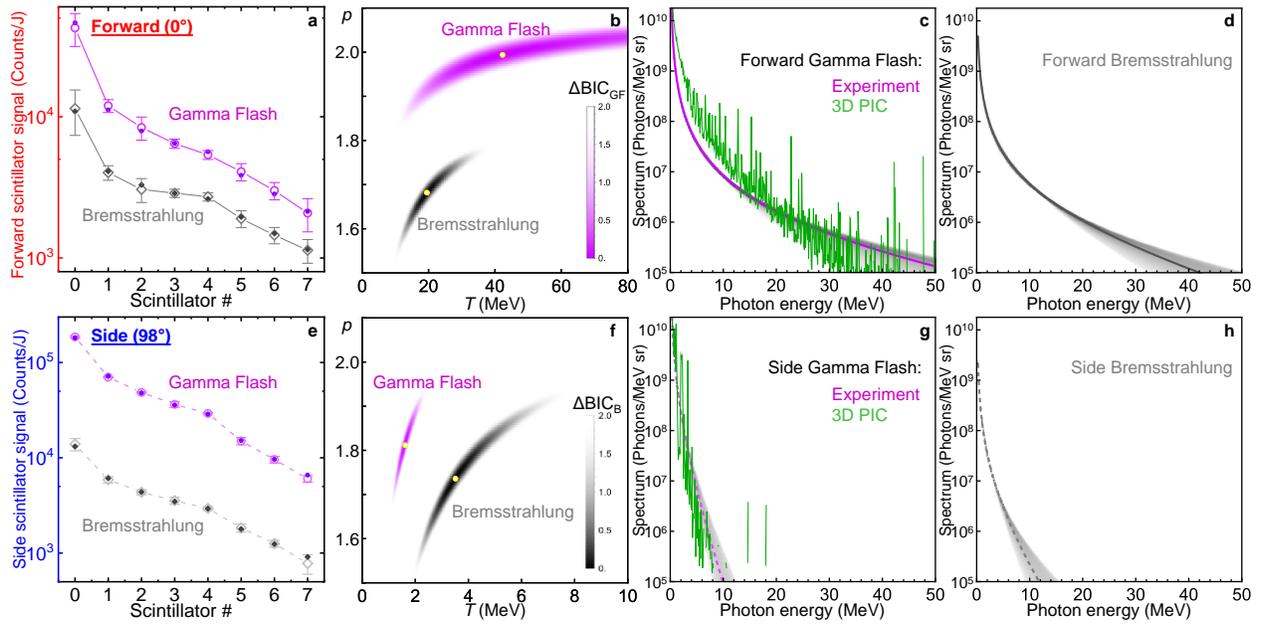

**Fig. 2. Gamma Flash and Bremsstrahlung in the experiment.** Top row: forward (0°, laser propagation direction) spectrometer, bottom row: side (98°) spectrometer, **Extended Data Fig. 3a**. **a**, **e** Scintillator signal vs. scintillator number; open circles are the intercepts of the fitting curves (**Extended Data Fig. 4** and **Extended Data Fig. 5**) corresponding to the surface process, Gamma Flash; open rhombi are the slopes (here multiplied by 5 µm) corresponding to the bulk process, Bremsstrahlung. The error bars (some are hard to see) are the fit errors. The small solid circles and rhombi are reconstructed scintillator signals assuming a power-law with cut-off spectral shapes with number of photons $N_p$, power-law parameter $p$ and cutoff temperature $T$: $dN/d\mathcal{E} = (N_p/T)(T/\mathcal{E})^p \exp(-\mathcal{E}/T)$ (Methods). **b**, **f**, Two-parameter maps ($p$, $T$) where the yellow circles indicate the best-fit parameters, while the colourscales (same for both frames) show the Bayesian Information Criterion (BIC) difference, ΔBIC, from 0 (the best fit) to 2. **c**, **g**: Gamma Flash spectra reconstructed from the experiment (magenta line with shaded area showing reconstruction error) and 3D PIC simulation (green). **d**, **h** Bremsstrahlung spectra reconstructed from the experiment.

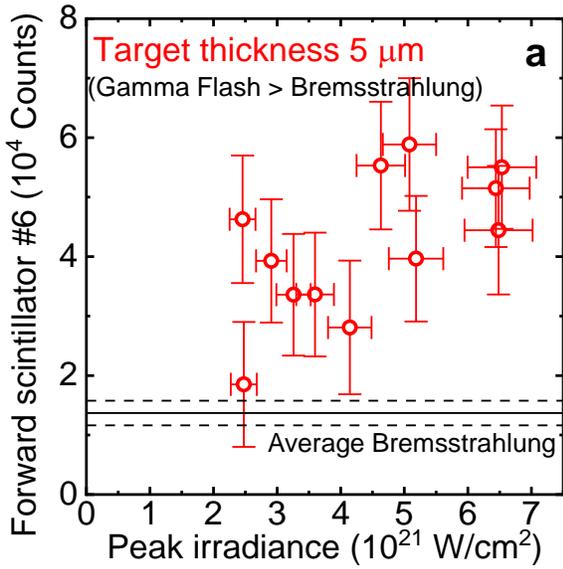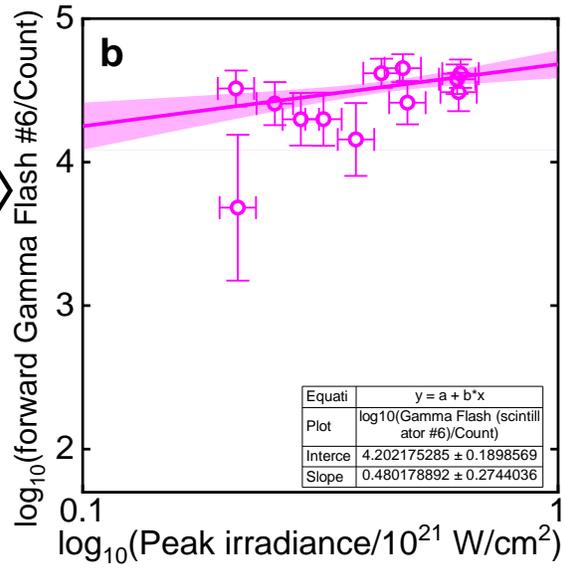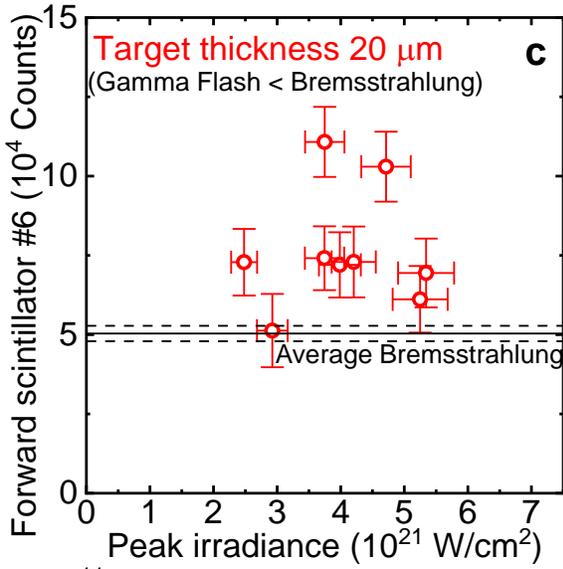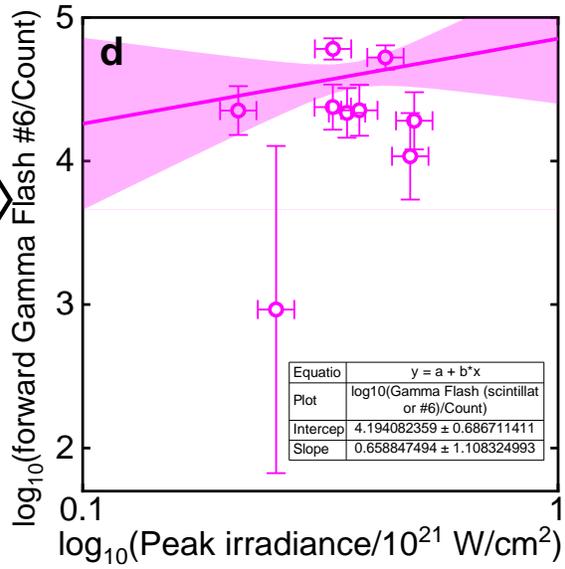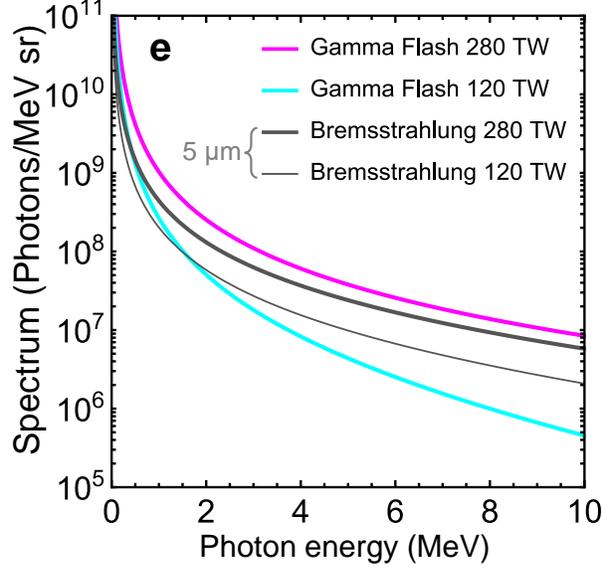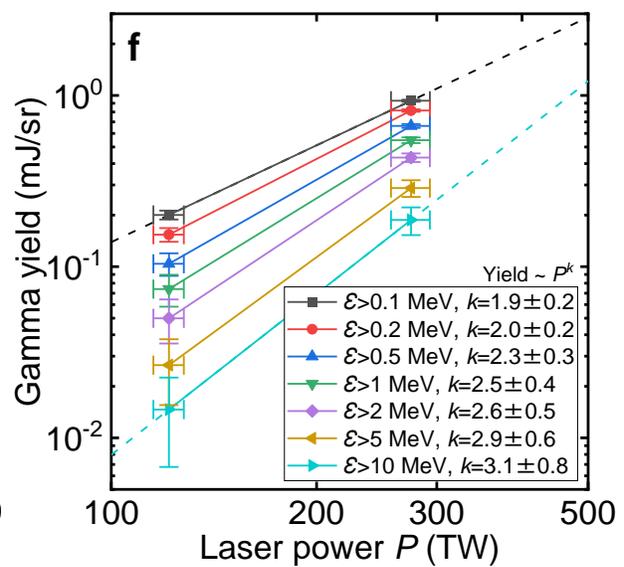

**Fig. 3. Experimental Gamma Flash scalings. a**, **c**, High-energy photon yield dependences on peak laser irradiance estimated from on-shot data. **b**, **d**, Gamma Flash photon yield irradiance scalings derived from the total photon yield (a, c) by subtracting the average Bremsstrahlung contribution (solid lines in (a, c); dashed lines indicate standard errors). Gamma Flash dominates over Bremsstrahlung for the thin (5 μm) target (a, b), while for the thick (20 μm) target Bremsstrahlung is stronger (c, d). The laser pulse energy and power were approximately constant during all the shots, **Extended Data Fig. 6c,d**; the irradiance difference stemmed from shot-to-shot spot quality variations. Error bars in (b, d) result from error propagation and include contributions from the raw data errors (a, c) and linear fit intercept and slope errors, **Extended Data Fig. 4**, used for the Bremsstrahlung subtraction. **e**, Gamma Flash and Bremsstrahlung spectra at two laser power values. **f**, Gamma Flash yield dependence on laser peak power with two experimental points for 122±6 and 280±20 TW; horizontal error bars are standard deviations, vertical error bars are propagated from the reconstructed experimental spectra. The legend shows colour-encoded photon energy ranges and scaling exponents of the corresponding power-law fits.

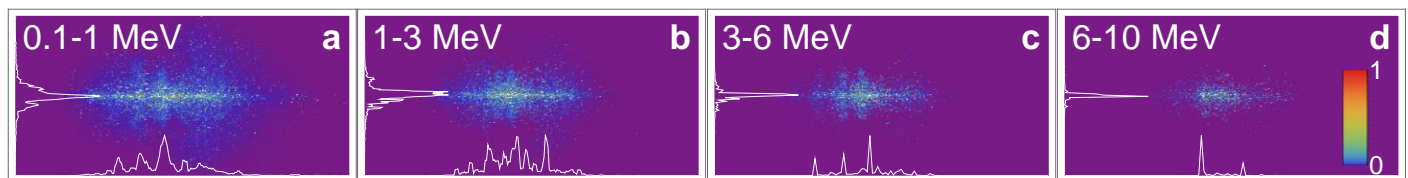
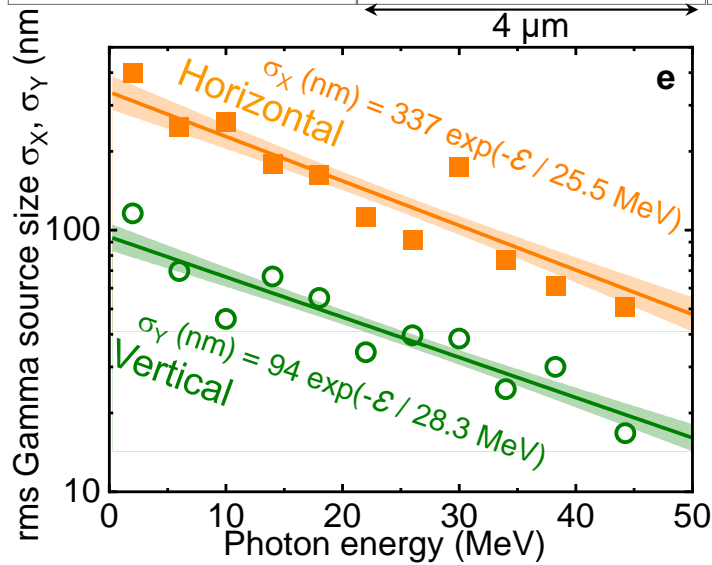
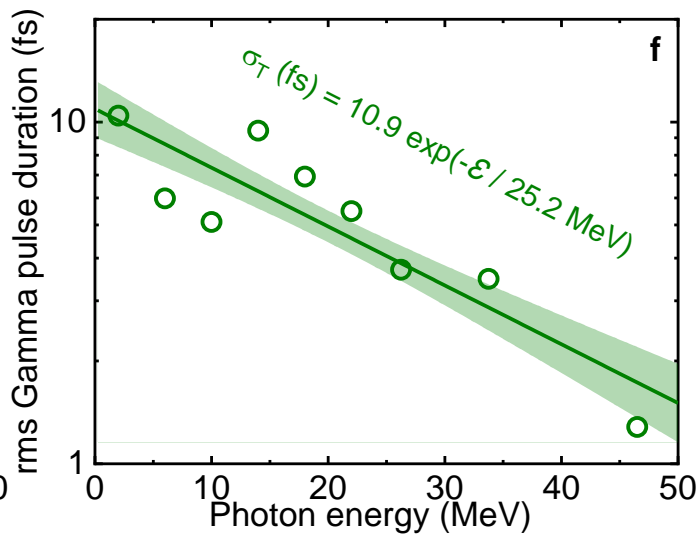
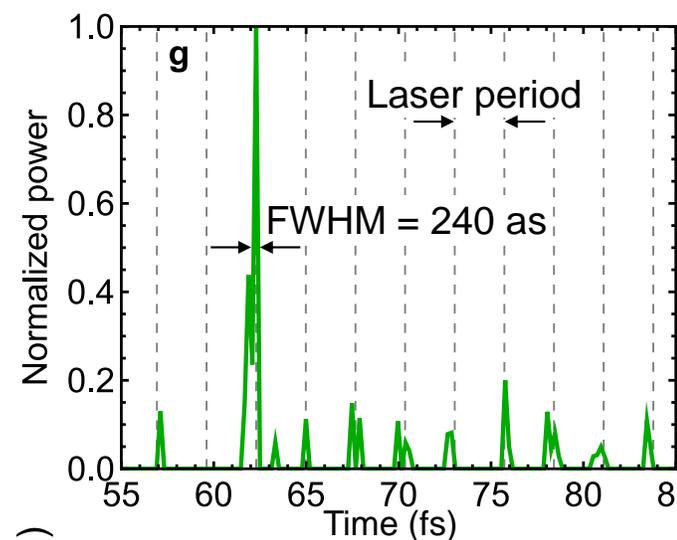
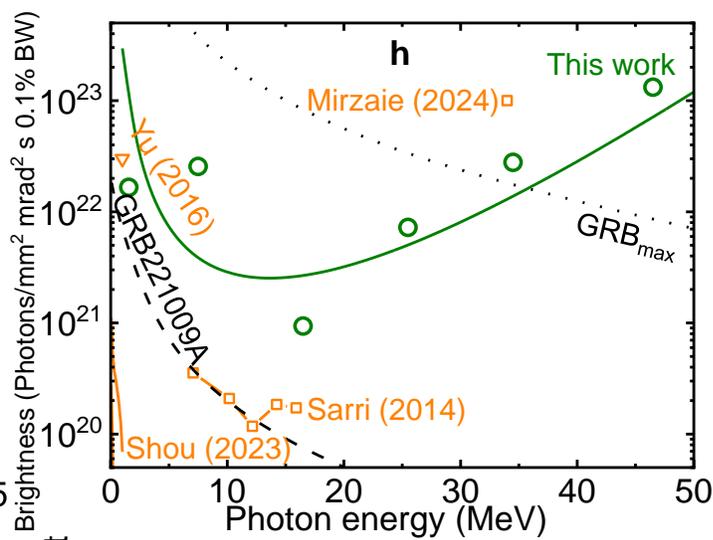
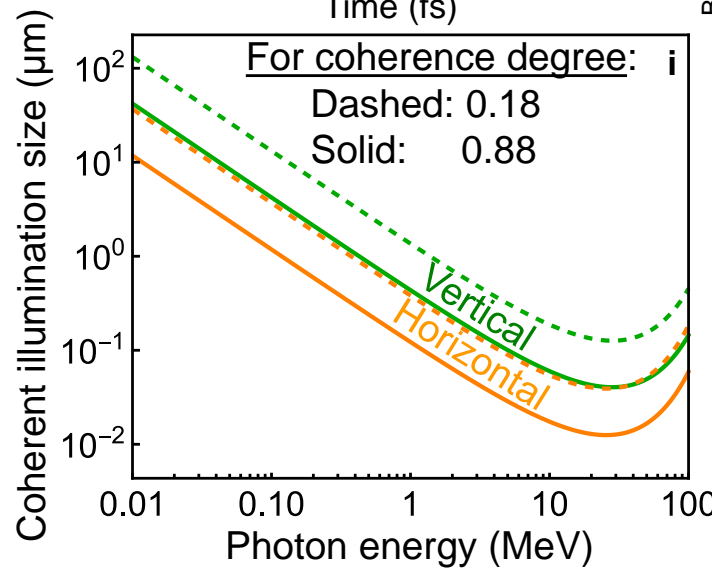
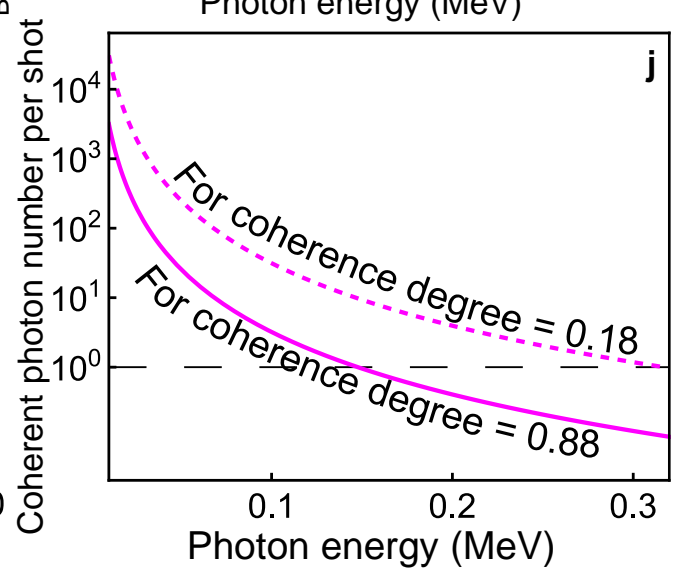

**Fig. 4. Gamma Flash source size, temporal shape, brightness, and spatial coherence. a-g**, 3D PIC simulation. **a-d**, The gamma source spatial distribution, 2°×2° forward emission with specified photon energies; each frame is 4 μm × 2 μm. **e**, **f**, Gamma Flash rms source size and rms pulse duration vs. photon energy: exponential fits (lines) with 67% confidence bands (shaded regions). **g**, Attosecond gamma pulse train for the 2°×2° forward emission of 20-25 MeV photons. **h**, Gamma Flash brightness. Green circles and fit: brightness calculated from the 3D PIC simulation. Orange lines and points: brightness in other experiments[5,98-100]. Black dashed and dotted lines: estimated brightness of the strongest Gamma Ray Burst ever detected, GRB221009A, assuming 10 and 0.3 light second source size, respectively (Methods). **i**, Coherent illumination size[101] estimated as $0.16\lambda L/\rho$ (solid) and $0.5\lambda L/\rho$ (dashed) for two values of the required coherence degree, where the wavelength $\lambda \approx 1.24$ pm/$\hbar\omega$[MeV], propagation distance $L = 10$ cm; the source radius $\rho$ is estimated from fits (e). **j**, Coherent photon number per shot estimated from the coherent illumination size and experimental Gamma Flash spectrum, **Fig. 2c**, assuming 50% bandwidth.

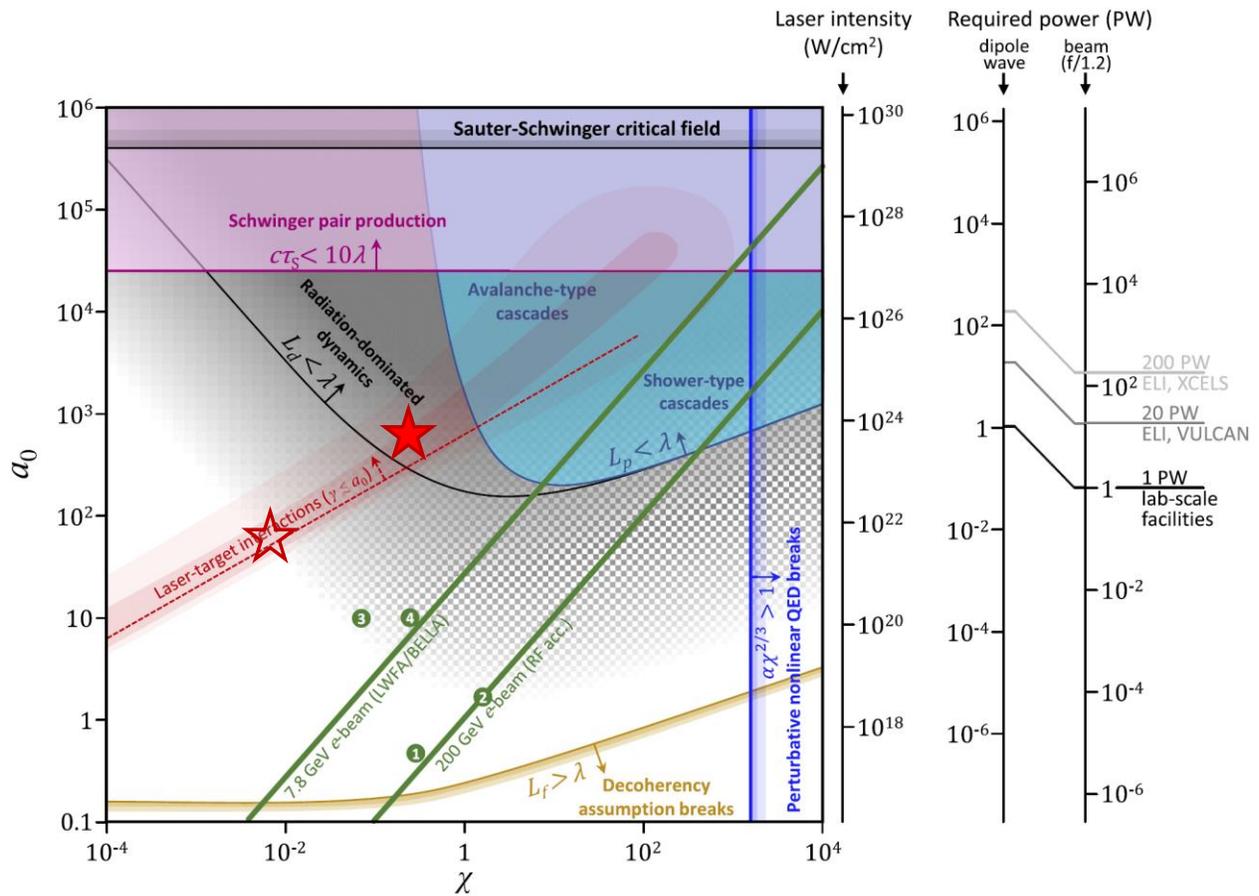

**Fig. 5.** Interaction regimes in the plane of quantum nonlinearity parameter χ and dimensionless field amplitude $a_0$, the original figure adopted from[2] with permission. The open star denotes a would-be regime corresponding to average $\chi_e$ over the entire focal volume and original laser irradiance. The solid star denotes our regime characterized by the peak irradiance and maximum $\chi_e$ parameter achieved in the overlap of the reflected intensified field with the electron density peak (fold singularity, **Extended Data Fig. 2d**).

**Extended figures**

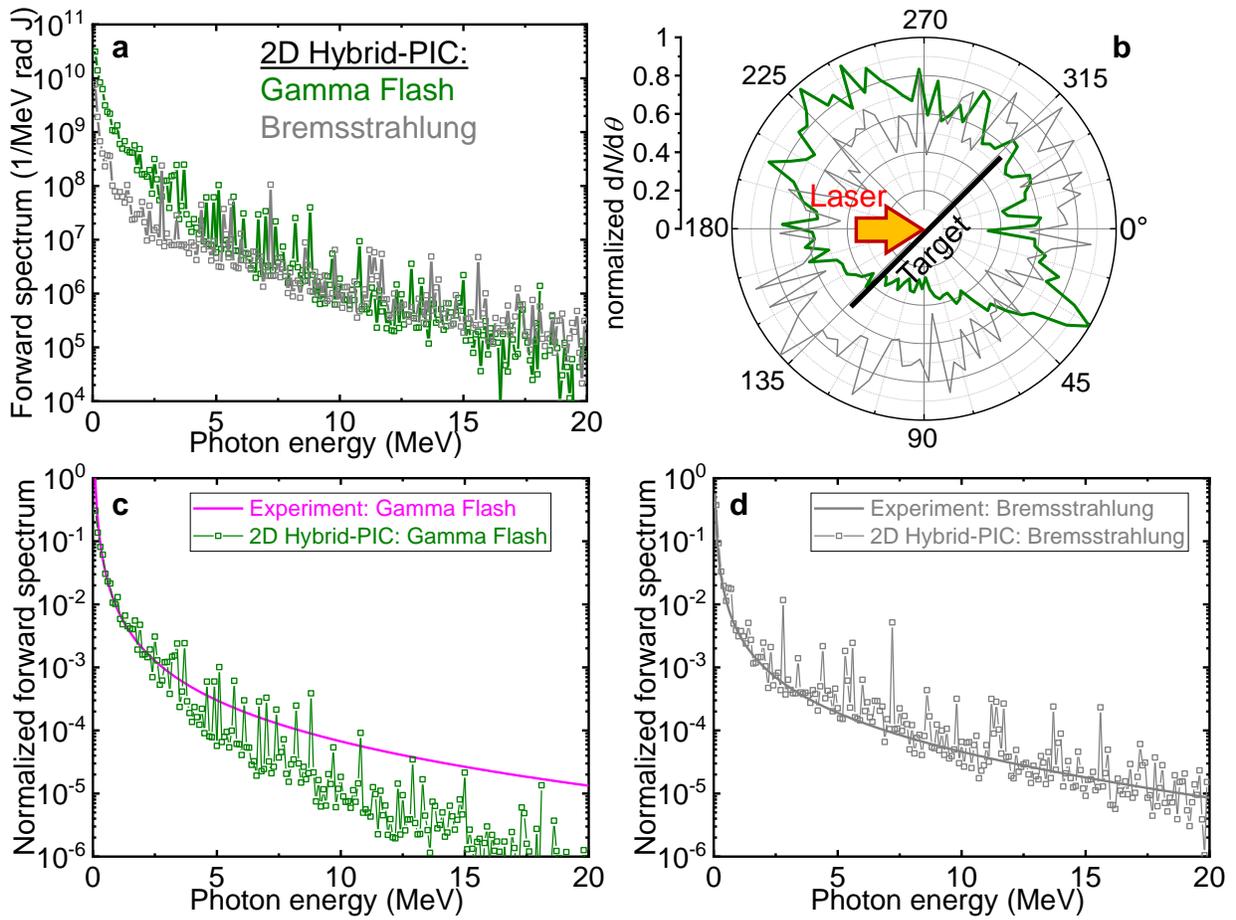

**Extended Data Fig. 1.** Gamma Flash and Bremsstrahlung in the 2D Hybrid-PIC simulation. **a**, Gamma Flash and Bremsstrahlung forward spectra normalized by the (same) laser energy. **b**, Normalized angular distributions at 0.4 MeV photon energy. **c**, **d** Comparison of 2D Hybrid-PIC simulation with the experiment: normalized forward spectra of Gamma Flash (c) and Bremsstrahlung (d).

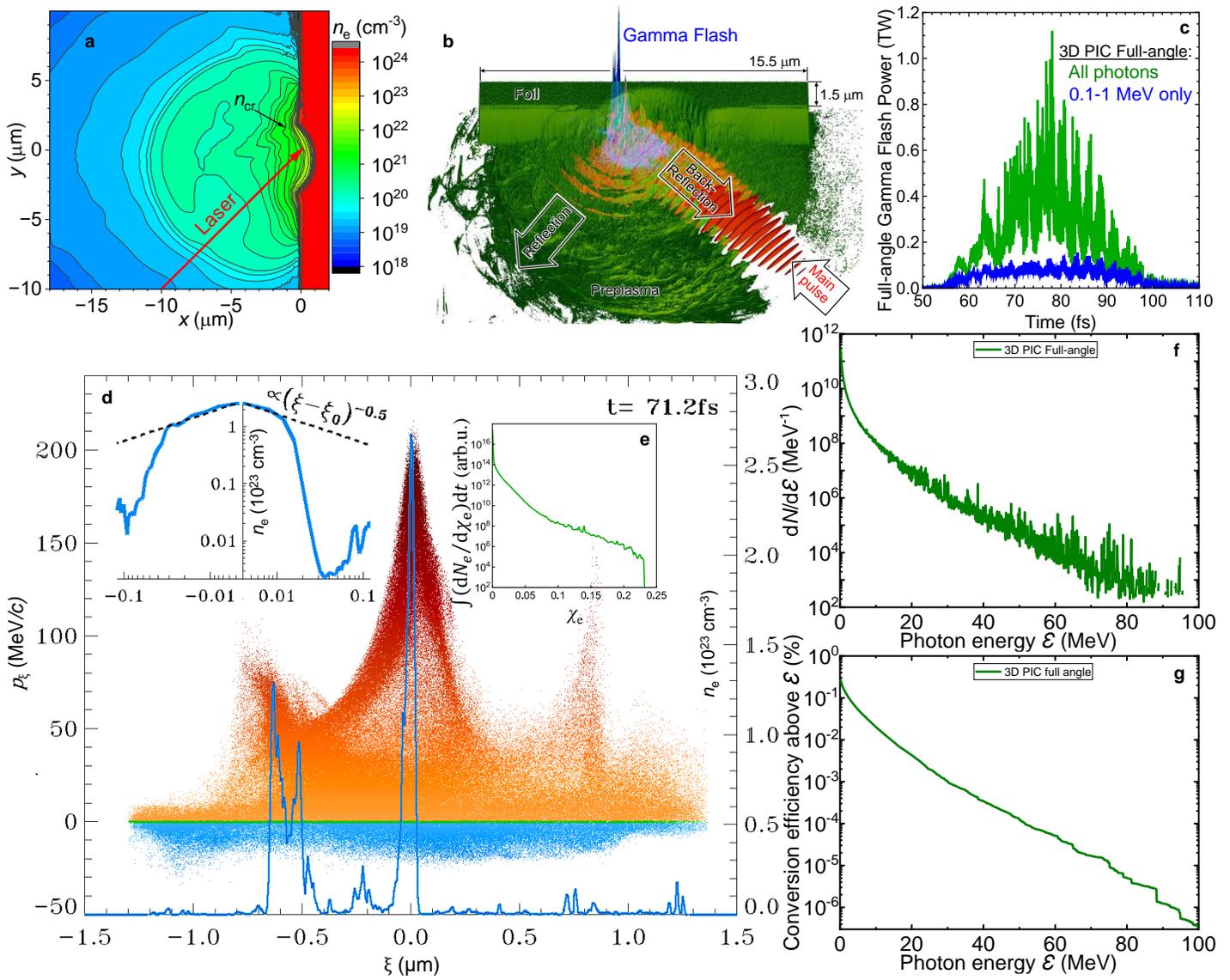

**Extended Data Fig. 2.** **3D hydrodynamic and 3D PIC simulations. a**, 3D hydrodynamic simulation of preplasma used in both 2D Hybrid-PIC and 3D PIC simulations. **b**-**g**, 3D PIC simulation results. **b,** The electron density (green) and the electromagnetic field energy density (red) at 4 fs before the main pulse centre reaches the unperturbed foil surface. The accumulated energy density of gamma photons (blue) with photon energy above 0.1 MeV at the end of Gamma Flash. **c**, Angularly-integrated (4π sr) Gamma Flash power. **d**, Left axis: electron phase space ($\xi, p_\xi$) in preplasma along the specular direction $\xi$; the blue lineout shows the electron density (right axis), its central part is shown in the inset in the log-log scale demonstrating $\propto(\xi-\xi_0)^{-1/2}$ dependence characteristic for the fold singularity. **e**, Time-integrated distribution of the quantum nonlinearity parameter $\chi_e$ in a 2-µm cube around the laser focus, where practically all the gamma photons are generated. **f,g** Angularly-integrated Gamma Flash spectrum and conversion efficiency.

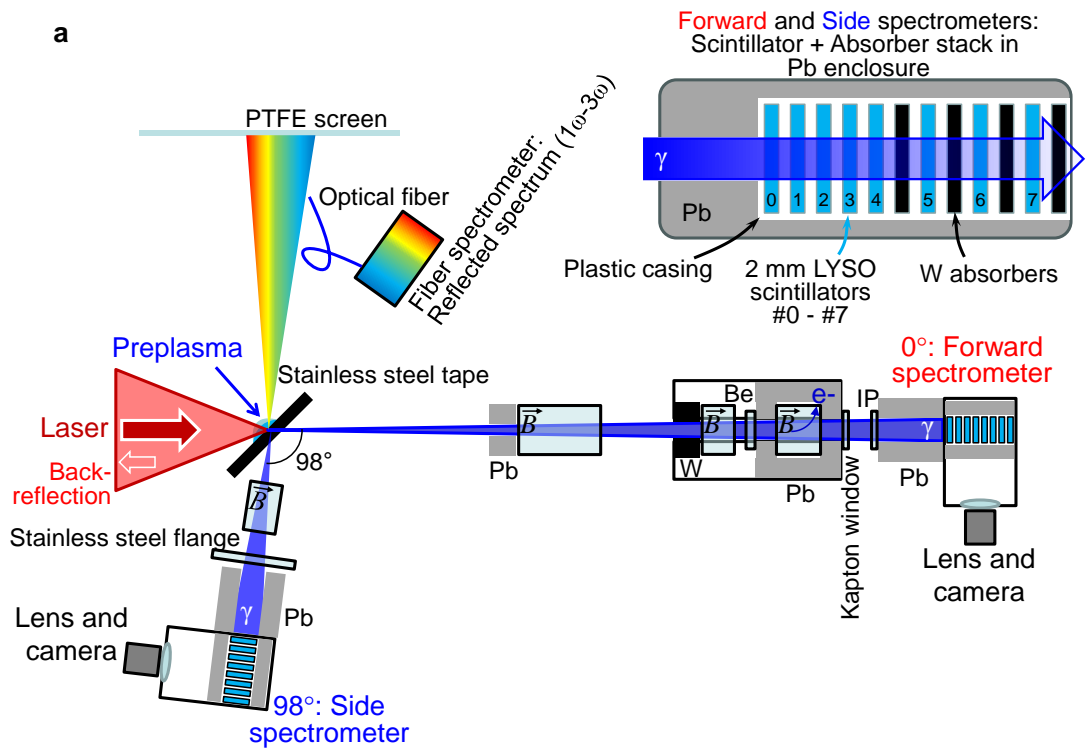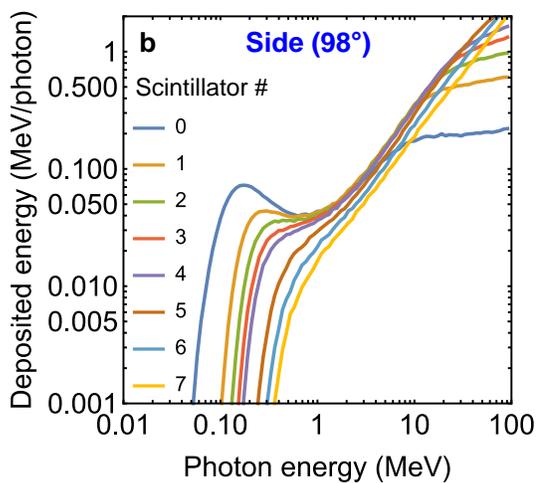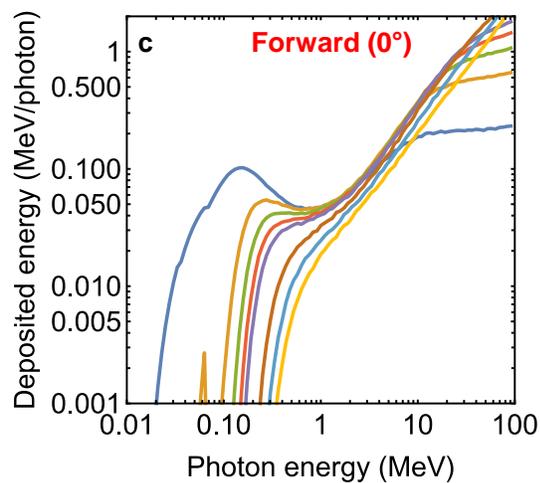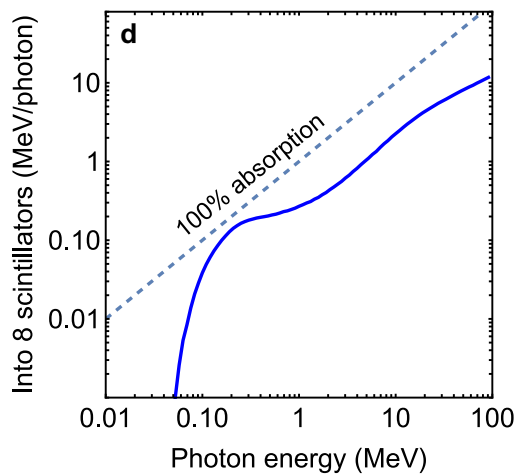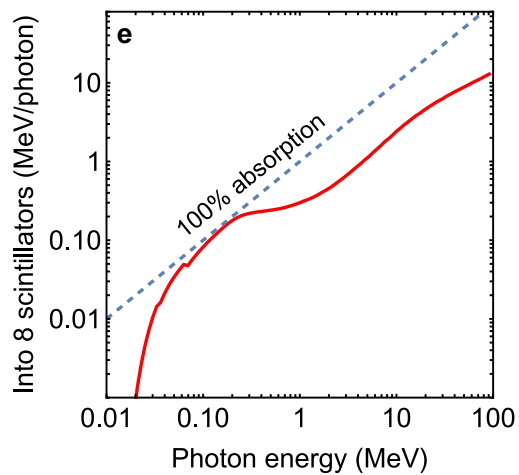

**Extended Data Fig. 3.** **Experimental setup. a**, The p-polarized (in the figure plane) laser pulse is focused onto the surface of a 20 mm wide stainless-steel tape at 45° incidence. The tape thickness is varied from 5 to 20 µm with 5 µm steps. The preplasma is tailored by controlling the laser prepulse. The interaction is diagnosed by measuring back-reflected energy and specularly reflected spectrally-resolved 1$^{st}$, 2$^{nd}$, and 3$^{rd}$ harmonics, ensuring the same conditions on the target surface, **Fig. 1d**. The generated high-energy photons (sub-MeV to tens of MeV photon energies) are measured with forward (0°, along the laser axis) and side (98° to the laser axis) linear absorption spectrometers consisting of 2 mm thick LYSO scintillators and tungsten absorbers. The high-energy photons going through the scintillators cause emission of ~400-500 nm light imaged by lenses onto cameras. To avoid excessive noise, the electrons and ions are deflected by magnets (denoted as rectangles with $\vec{B}$); the spectrometers are additionally shielded with led (Pb) and tungsten (W) collimators and blocks. **b, c**, Scintillator response, deposited energy per photon simulated by the GEANT4 code[73] including losses and secondary particle generation in the beamlines and spectrometers (Methods). **d, e**, Sum of deposited energies from (b, c) into the eight scintillators.

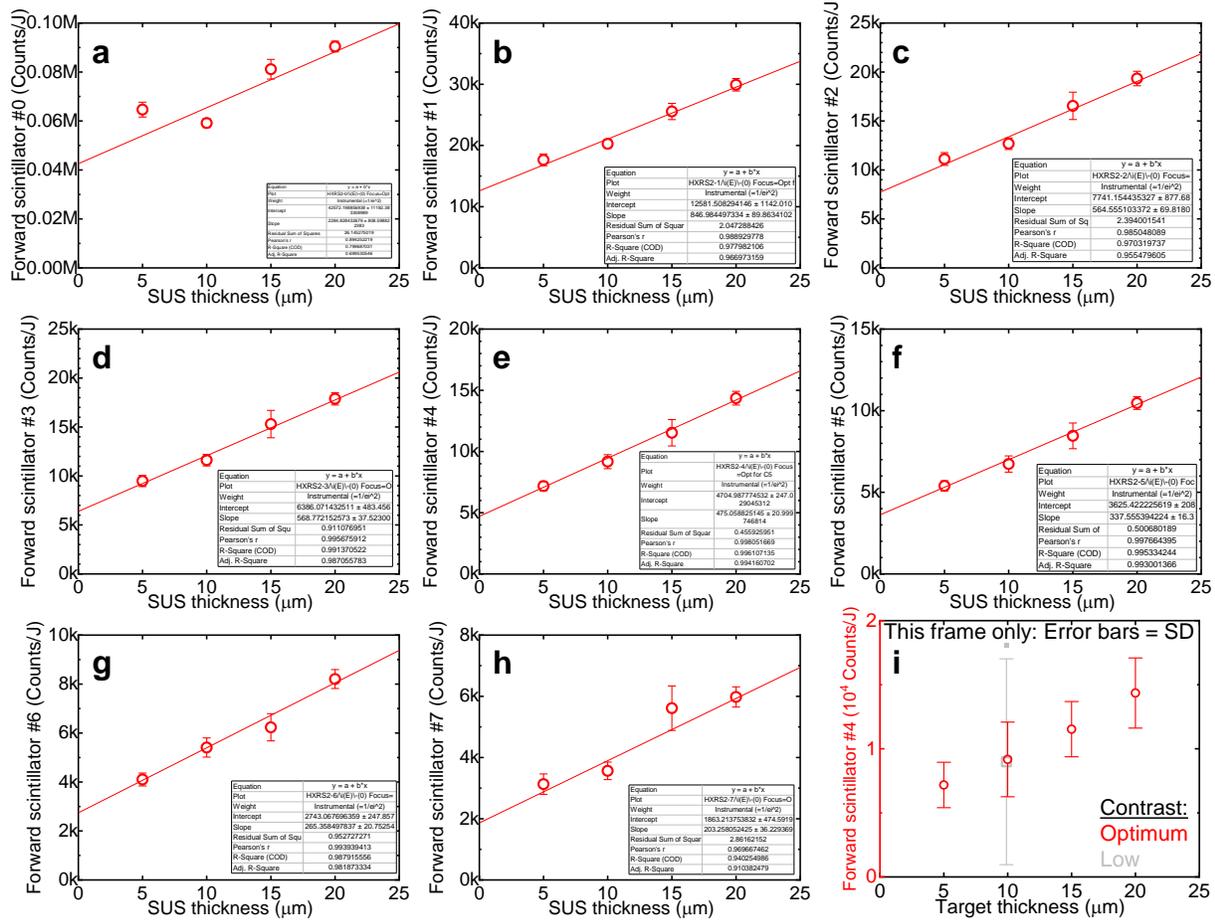

**Extended Data Fig. 4.** **Experiment: forward spectrometer data.** Summary of all performed high-irradiance shots (22 for 5 µm, 26 for 10 µm, 4 for 15 µm, and 24 for 20 µm target thickness, 76 shots total). **a-h**, Experimental dependences of all scintillator photon yields on target thickness, the error bars are standard errors of shot-to-shot fluctuations; the lines are linear fits. **i**, the same as (e) but the error bars are the standard deviations; the grey points show a lower-contrast experiment attempt: the open square is the average with standard deviation, the solid square is the maximum.

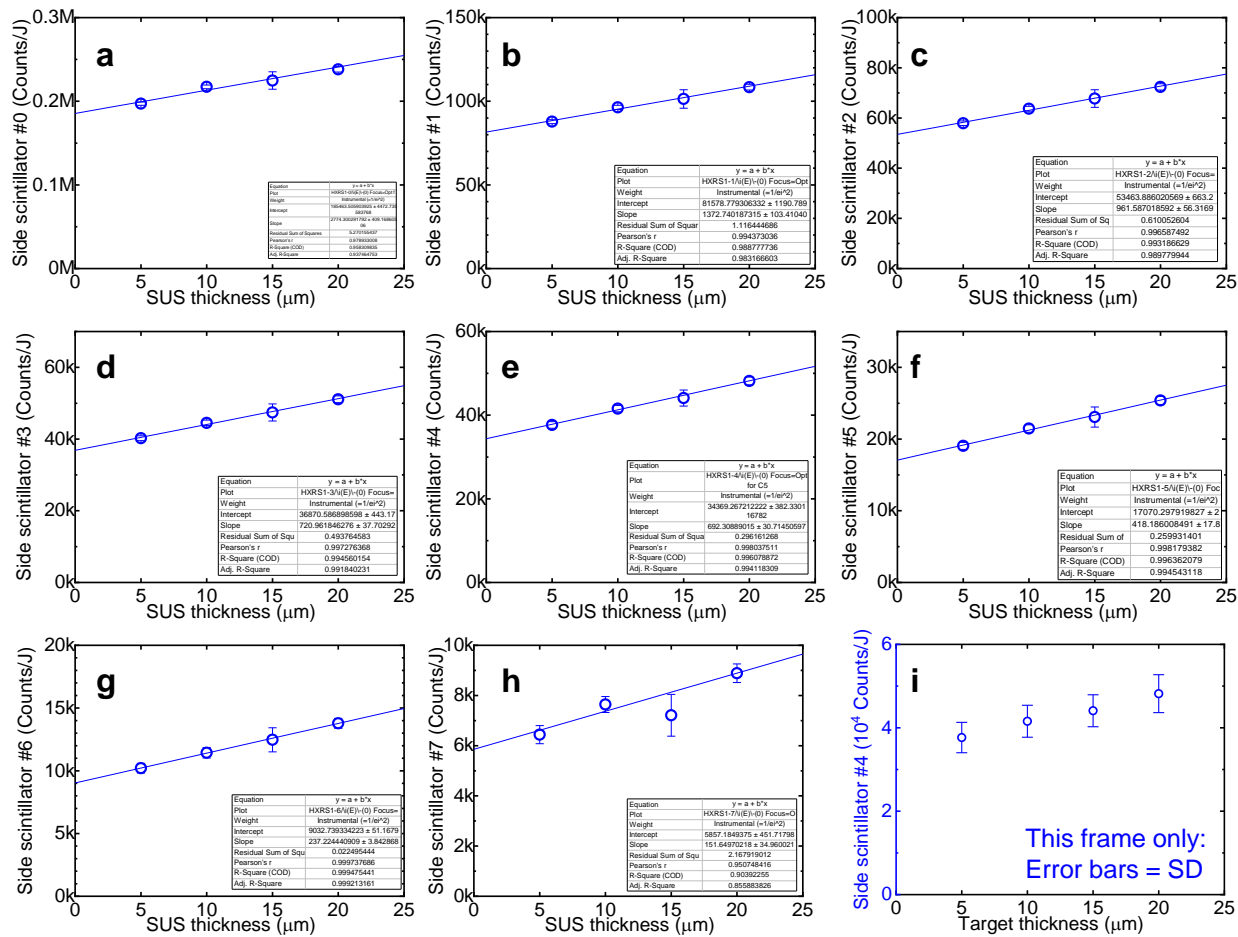

**Extended Data Fig. 5.** **The same as Extended Data Fig. 4 for the side spectrometer.** The data from the same shots as Extended Data Fig. 4, however, this spectrometer missed one shot due to the plasma-generated electromagnetic pulse interference[102]; thus, a 21-shot average is shown for the 5 μm target instead of 22 shots in the forward spectrometer data.

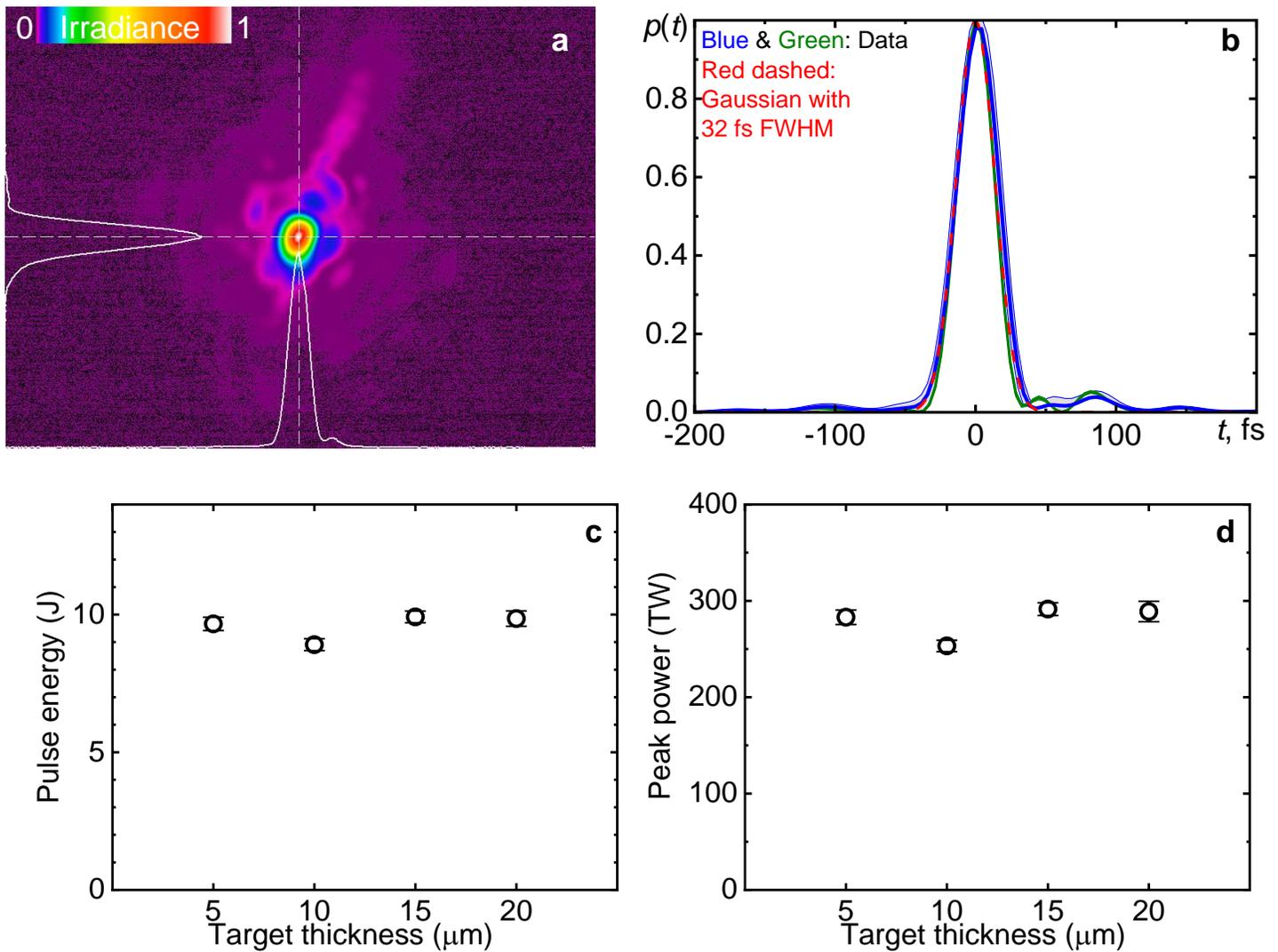

**Extended Data Fig. 6.** **Laser properties in the experiment. a**, Focal spot shape measured before the experiment with ~10% power mode. The full frame size is 41.3 μm × 31.0 μm. **b**, Temporal pulse shapes measured before the experiment with ~10% power mode (blue) and calculated from the measured full-power on-shot spectra and measured average spectral phase (green). The shaded regions show standard deviations of the shot-to-shot fluctuations. **c, d**, Laser pulse energy and peak power during shots for each target thickness in the experiment. The error bars are standard deviations of shot-to-shot variations.

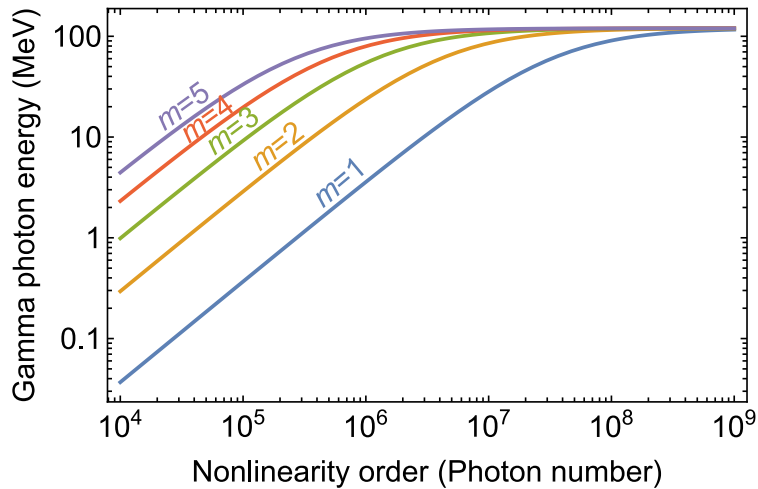

**Extended Data Fig. 7.** **Estimated maximum gamma photon energy vs. number of optical photons participating in single scattering event.** Different curves show estimates for the laser frequency ($m$ = 1) and its harmonics ($m$ = 2 to 5).